\documentclass{jfm}
\usepackage{graphicx}
\usepackage{epstopdf, epsfig}
\usepackage{gensymb}
\usepackage{amssymb}

\makeatletter
\newcommand{\mypm}{\mathbin{\mathpalette\@mypm\relax}}
\newcommand{\@mypm}[2]{\ooalign{%
  \raisebox{.1\height}{$#1+$}\cr
  \smash{\raisebox{-.6\height}{$#1-$}}\cr}}
\makeatother

\shorttitle{Triggering flow asymmetry by lamella deflection}
\shortauthor{K. Regulagadda, S. Bakshi and S. K. Das}

\title{Triggering flow asymmetry by lamella deflection during drop impact on superhydrophobic surfaces}

\author{K. Regulagadda\aff{1},
  S. Bakshi\aff{1}
  \corresp{\email{shamit@iitm.ac.in}}
 \and S. K. Das\aff{1,2}}

\affiliation{\aff{1}Department of Mechanical Engineering, Indian Institute of Technology, Madras 600036, Tamil Nadu, India
\aff{2}Current Address: Department of Mechanical Engineering, Indian Institute of Technology, Ropar 140001, Punjab, India}

\begin{document}

\maketitle

\begin{abstract}
Water drop impacting a superhydrophobic surface (SHS) rebounds completely with remarkable elasticity. For such an impact, the balance between the inertial and capillary forces ascertain the contact time. This is found to be fairly constant for any macroscopically flat SHS and for a given drop volume. Recently, various studies have shown that breaking the radial symmetry during the drop impact can significantly reduce the contact time below that of a flat SHS. One such study has been performed on a cylindrical SHS with a curvature comparable to the drop. The reduction in contact time has been attributed to the radially anisotropic flow imparted by the tangential component of momentum and the elliptical footprint of the drop during the crash. Here, we perform drop impact experiments on bathtub-like SHS and show that the radial anisotropy can be triggered even in the absence of both the criteria mentioned above. This is shown to be a consequence of the lamella deflection during the spreading of the drop. The reduction in contact time is quite clearly evident in this experimental regime.  
\end{abstract}

\section{Introduction}
Droplet impinging onto a solid surface is ubiquitous in nature and practical applications and is being studied by researchers and scientists for more than a century. The impact patterns discussed in the pioneering work of \cite{Worthington01011876} established a significant understanding of the hydrodynamics of impact. Various studies over the years revealed that wettability of the surface is a key factor in deciding the impact outcome.  The first investigation of \cite{Wenzel1936} in designing water-repellent fabrics is a significant progress in analysing the surface wettability. This analysis is extended by \cite{Cassie1944} to porous surfaces. \cite{Shibuichi1996} have shown that the manipulation of surface texture can produce remarkable water-repellency which are now known to be superhydrophobic surfaces (SHS).

Spray cooling is a technique where the drops interact with the surface and take away a significant amount of heat because of the phase change during the impact. However, when the surface temperature is at/above the Leidenfrost point, complete rebound can be observed \citep{Chandra13, Tran2012} because of the formation of stable vapour film between the impacting drop and the surface. This drastically reduces the efficiency of spray cooling process since the interaction of drop with the surface is reduced by the non-conducting vapour film in between. On the contrary, some applications like anti-icing \citep{Cao2009, Jung2011}, self-cleaning \citep{Barthlott1997, Marmur2004, Quere2008} etc. require complete lift-off of the drop from the surface during impact which is observed on low energy surfaces like SHS (without heating the surface). Rapid shedding of drops makes these surfaces ideal for anti-icing applications. \cite{Richard2002} showed that the contact time of a bouncing drop on a macroscopically flat SHS is constant for a wide range of Weber numbers ($We > 1$) where $We = \rho U_0^2R_0/\sigma$ ($\rho$ is the density of the liquid, $U_0$ is the impact velocity of the drop, $R_0$ is the radius of the drop before impact and $\sigma$ is the surface tension of the liquid). The phenomenon is explained by the balance of inertia and surface tension forces giving rise to inertio-capillary time scale $\tau = \sqrt{\rho R_0^3/\sigma}$. The contact time in the above mentioned $We$ regime is found to be around 2.6$\tau$, which can be thought of a theoretical limit (minima) for the contact time on any flat surface. During the impact, the drop forms a pancake structure at maximum spread and recoils to lift-off maintaining radial symmetry all throughout.

Ice accretion on surfaces takes place if the time scale of ice nucleation matches with the contact time of the drop during a freezing rain. Despite exhibiting a complete drop rebound, the SHS still suffer from ice formation under adverse environmental conditions \citep{Kulinich2009, Meuler2010, Jung2012}. The theoretical limit to the contact time of drop impact on a flat SHS thus poses a severe limitation on the anti-icing application of the surface. However, recent studies \citep{Bird2013, Liu2014, Liu2015a, Weisensee2016} have shown that it is possible to reduce the contact time by triggering radially anisotropic flow by breaking the radial symmetry of the impact. \cite{Bird2013} introduced macroscopic ridges on a flat SHS  while \cite{Liu2015a} performed experiments on cylindrical SHS with curvature comparable to the drop. The main reasons for flow anisotropy leading to rapid bouncing were identified as the elliptical footprint of the drop on the substrate and the tangential component (parallel to the surface) of momentum during the drop impact. Recently, \cite{kartik2017} showed that the post-impact morphology of this anisotropic flow could be very different based on the impact configurations.

\begin{figure}
  \centerline{\includegraphics[width = \linewidth]{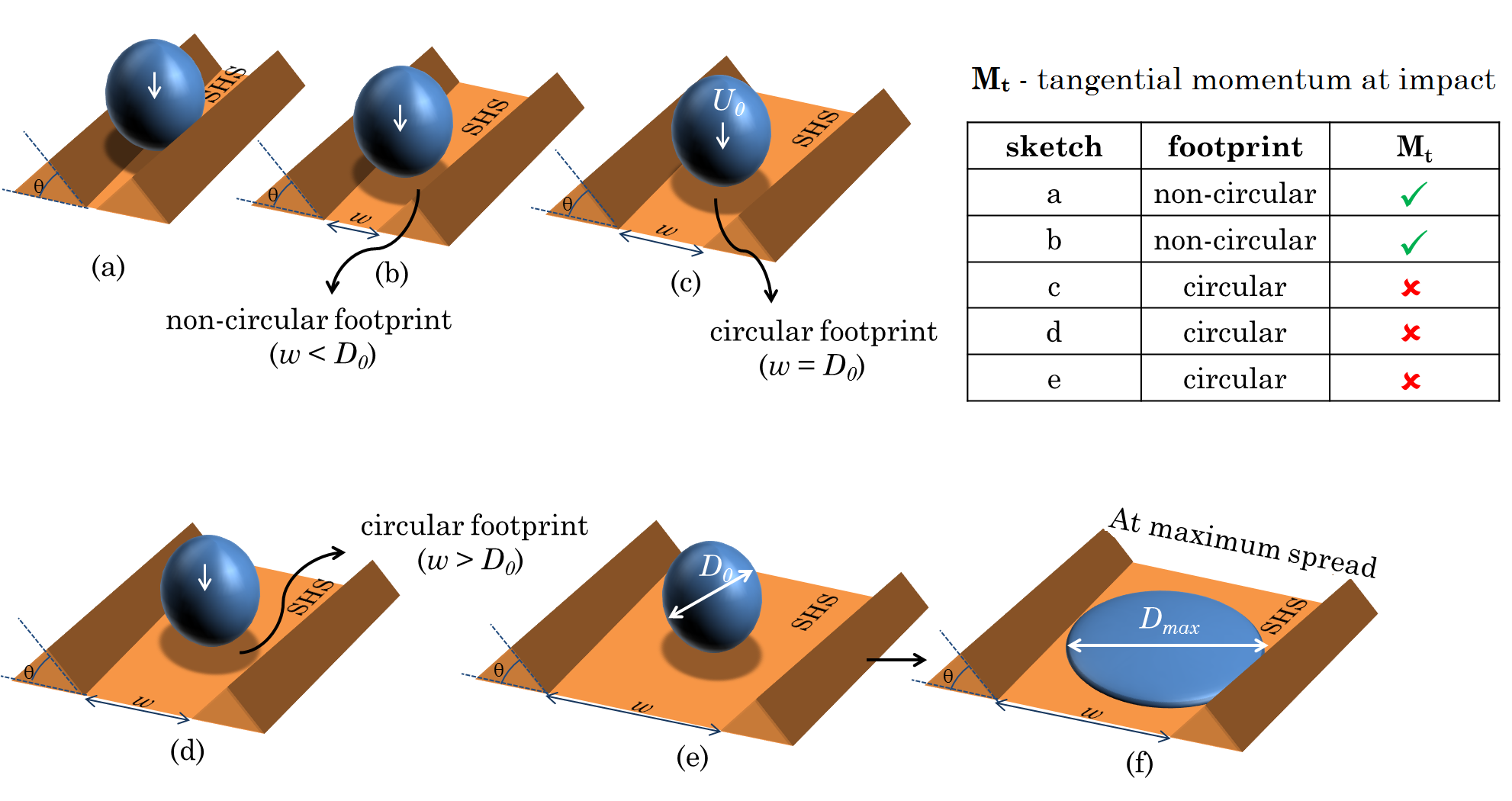}}% Images in 100% size
  \caption{Sketch showing the configuration of impact. Sketches \textbf{a} and \textbf{b} indicate the configuration where $w < D_0$. It can be clearly seen that the footprint or the shadow of the drop on the substrate intersects with the inclined faces. Thus, tangential component of momentum at impact exists along with a non-circular footprint. However, sketches from \textbf{c}-\textbf{f} indicate the absence of both where $w \geqslant D_0$. Sketch \textbf{f} indicates the maximum $w$ considered in the experiments so that the liquid lamella never climbs the inclined face.}
\label{substrate}
\end{figure} 

In the present work, we have designed our experiments so that the drop footprint can be changed from non-circular to circular. Consequently, the tangential component of incident velocity changes from non-zero to zero. The detailed description of experimental setup and procedure is given in \S\ref{EM}. From these experiments, we can observe an intermediate regime where the radially anisotropic flow is triggered even in the absence of any tangential component of velocity or non-circular footprint. The reduction in contact time can be observed in this regime as well. 

\section{Experimental Method}\label{EM}
Isosceles trapezoidal cross-sections (bathtub-like) are machined on substrates (see Figure \ref{substrate}) using wire-cut Electro Discharge Machining process with varying flat base width (0 $\leqslant w \leqslant$ 8 mm). When $0 < w < D_0$ ($D_0$ is drop diameter), it should be noted that the drop impinges partly onto the inclined face during the crash (see Figure \ref{substrate}a and b). Furthermore, the footprint is non-circular. To eliminate these two effects on the impact hydrodynamics, the width of the flat region ($w$) is gradually increased. Thus, $w \geqslant D_0$ indicates the configuration which has zero tangential momentum with a circular footprint (see Figure \ref{substrate}c-1f). The maximum value of $w$ is selected in such a way that the liquid remains within the flat valley at the maximum spread on the surface (see Figure \ref{substrate}f) during the impact for the selected $We$ which essentially means that this surface acts as a flat SHS. The angle ($\theta$) of the trapezoidal sections are selected to be 25$\degree$, 45$\degree$, and 60$\degree$. All the substrates are coated with superhydrophobic coatings from Ultra Ever Dry, Inc. The contact angle and the roll-off angle for the flat substrate is found to be more than $160\degree$ and less than $5\degree$ respectively, indicating the superhydrophobicity of the substrate \citep{Lafuma2003}. HPLC water droplets are impacted onto the center of the flat base with $We$ {\raise.17ex\hbox{$\scriptstyle\mathtt{\sim}$}} 21 from a calibrated needle with an outer diameter of 0.72 mm. The radius of the droplet in the experiments is $R_0 =$ 1.53 $\mypm$ 0.03 mm. The density and surface tension values of water are taken as 1000 kg/m$^3$ and 0.073 N/m respectively. The impacts are captured using synchronised high-speed cameras from the side and top views with a frame rate of 8000 Hz. All the images are processed using open source image software Fiji ImageJ(1.51n). The contact time is defined as the total interaction time for which the solid and liquid are in contact. The configuration of impact is represented as $S^{\theta}_{w/D_0}$. Thus, $S^{45\degree}_{0.17}$ represents the impact with $\theta = 45\degree$ and $w/D_0 = 0.17$. 

\begin{figure}
  \centerline{\includegraphics[width = \linewidth]{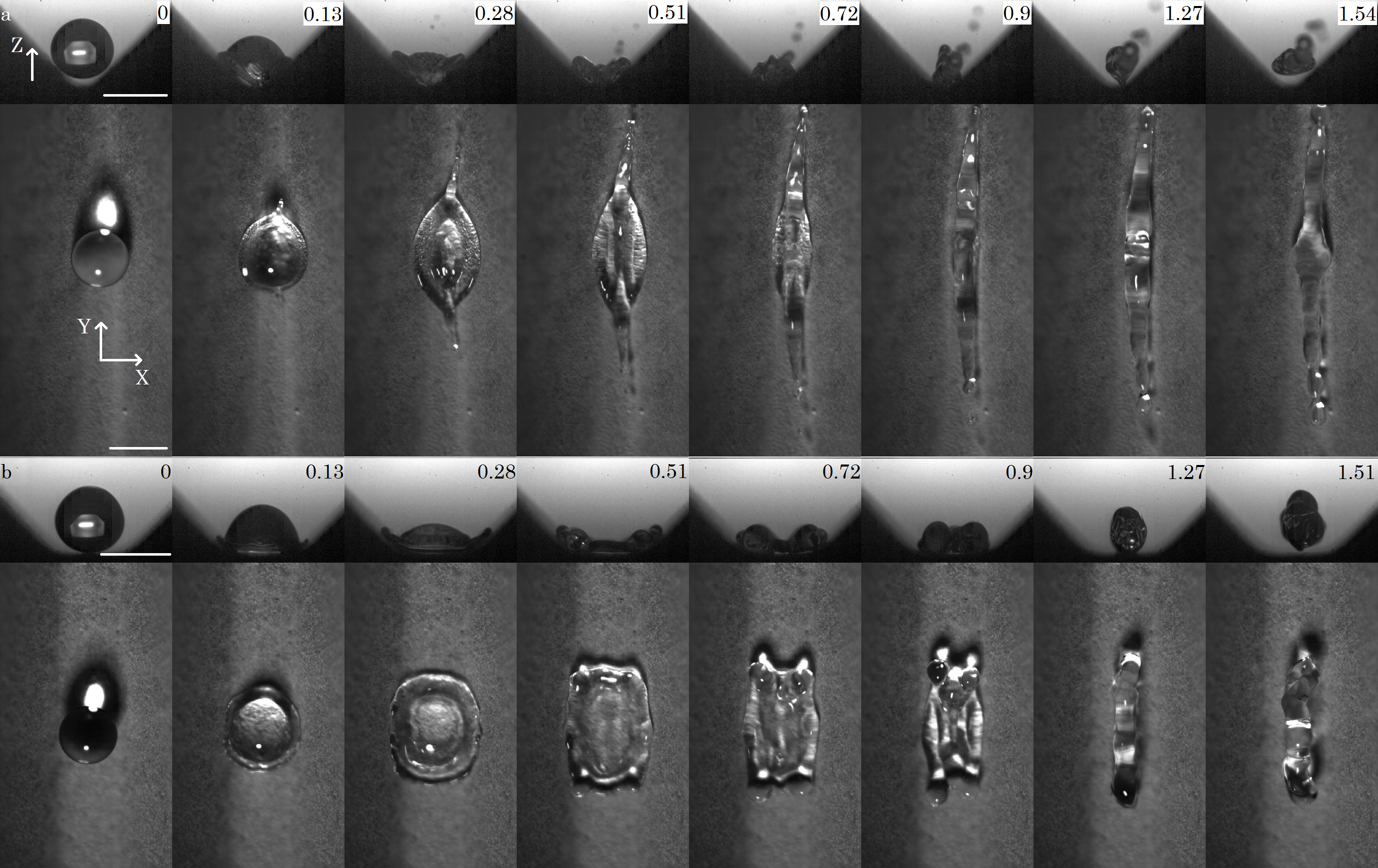}}% Images in 100% size
  \caption{ Side and top views of impact configuration (a) $S^{45\degree}_{0.17}$, $We$ = 19.8, (b) $S^{45\degree}_{1}$, $We$ = 22.2. The time indicated in each frame is non-dimensionalized with the inertio-capillary time scale ($\tau$). The scale bars in the insets represent 3 mm. The multimedia view is available in the supplementary material.}
\label{new_morphology}
\end{figure}

\section{Contact time}\label{CT}

Figure \ref{new_morphology} shows the morphology of drop impact for $S^{45\degree}_{0.17}$ and $S^{45\degree}_{1}$ configurations. For the cases where $w < D_0$, it is evident that the impact has a tangential component of momentum acting downhill which triggers the asymmetry and thus reduces the contact time \citep{Liu2015a, kartik2017}. This asymmetric momentum distribution produces a jet along the valley which can be noticed in the top view images of Figure \ref{new_morphology}a. The stretch of the jet in the valley is dependent on the magnitude of the tangential momentum imparted to the drop. As $w$ decreases, the spherical segment of the drop interacting with the inclined face increases producing higher tangential momentum \citep{kartik2017}. Secondary droplets are formed from the jet. It can be observed from the top view images that the drop forms a large central lobe with the jet ligaments on either side. Interestingly, the configurations where $w \geqslant D_0$ (see Figure \ref{new_morphology}b) produce asymmetric structures even though the tangential component of momentum is not present and the impact footprint is also circular. The recoiling of the drop is rapid downhill allowing early lift-off in comparison with a conventional flat SHS. Experiments were performed on the substrates with different $\theta$ values. Similar results were noticed for $S^{25\degree}$ and $S^{60\degree}$ (see supplementary material).

\begin{figure}
  \centerline{\includegraphics[scale = 0.3]{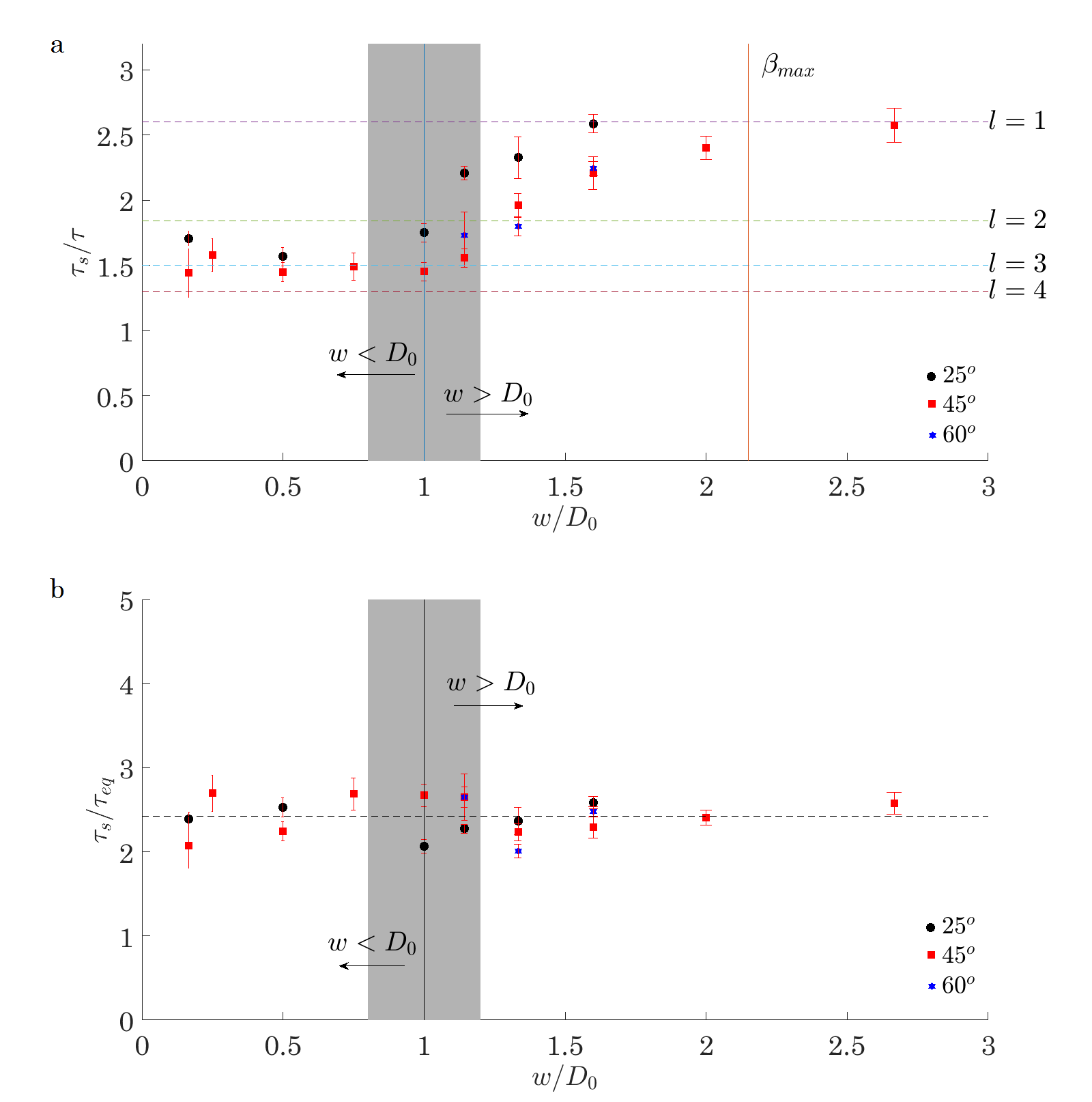}}
  \caption{\textbf{(a)} Plot showing the non-dimensional contact time ($\tau_s/ \tau$) with $w/D_0$ for $We$ {\raise.17ex\hbox{$\scriptstyle\mathtt{\sim}$}} 21. Here, $\beta_{max}$ indicates the maximum spread length non-dimensionalised with the drop diameter $D_0$ on a flat substrate for the selected $We$. The number the lobes formed during impact on macro-ridges as defined by \cite{Gauthier2015} is indicated by $l$. The dashed horizontal line acts as a guide for the eye. The error bars show the uncertainity in contact time. \textbf{(b)} Plot showing the variation of $\tau_{s}/\tau_{eq}$ with $w/D_0$. The data collapses onto a straight line indicating the inertio-capillary nature of the phenomenon.}
\label{contact_time}
\end{figure}

The contact times for all the configurations ($S_{w/D_0}^{\theta}$) at $We$ {\raise.17ex\hbox{$\scriptstyle\mathtt{\sim}$}} 21 are shown in Figure \ref{contact_time}a. For the regime where tangential component of momentum exists i.e., $w > D_0$, it can be observed that the contact time is approximately constant for all the configurations. However, when $w$ increases beyond $D_0$, the contact time rapidly changes with $w$ and eventually approaches the contact time on a flat surface ($S_{2.7}^{45 \degree}$ is equivalent to a flat SHS for the $D_0$ selected in the experiments). For, $w > D_0$, the asymmetry is triggered by the deflection of the lamella as the other mechanisms are absent here. With increase in the value of $w $ beyond $D_0$, the reduction in contact time is less. Finally, as $w$ {\raise.17ex\hbox{$\scriptstyle\mathtt{\sim}$}} $D_{max}$ (the maximum spread), the contact time is same as that of a flat SHS. 

Figure \ref{contact_time} also shows the contact time corresponding to the number of lobes as defined by \cite{Gauthier2015}. It has been explained in the literature that the number of lobes ($l$) formed just before lift-off determines the contact time of the drop during asymmetric bouncing \citep{Gauthier2015, kartik2017}. The asymmetry in the morphology results in redistribution of the volume of the droplet into lobes with smaller volumes. As the lobes have smaller volumes, the inertio-capillary time is also less and is given as $\sqrt{\rho V_l^3/\sigma}$  where $V_l$ is the volume of the lobes which is $(1/l)^{th}$ of the droplet volume ($V_0$). This approach holds good only when the volumes of the lobes are approximately equal. However, the volume distribution after the impact is not necessarily uniform, like in the present experiments. Thus we see in Figure \ref{contact_time} that the contact time falls in the window of non-integer values of $l$ (between 1-2 and 2-3 etc.). This is in contrast to the trend of variation of contact time with impact velocity on macro-ridges wherein a step-like behaviour is observed \citep{Gauthier2015}. Hence, we consider the largest lobe volume before lift-off to explain the contact time variation. We evaluated the largest lobe volumes in all the configurations and defined a $\tau_{eq} = \sqrt{\rho R_{eq}^3/ \sigma}$ where $R_{eq}$ is the equivalent radius of a sphere whose volume is equal to the largest lobe. The variation of $\tau_s/\tau_{eq}$ with $w/D_0$ is plotted in Figure \ref{contact_time}b. It can be observed that the data collapses onto a line with a value of $\tau_s/\tau_{eq}$ {\raise.17ex\hbox{$\scriptstyle\mathtt{\sim}$}} 2.4. This again highlights the inertio-capillary nature of the phenomenon.

\begin{figure}
  \centerline{\includegraphics[scale = 0.3]{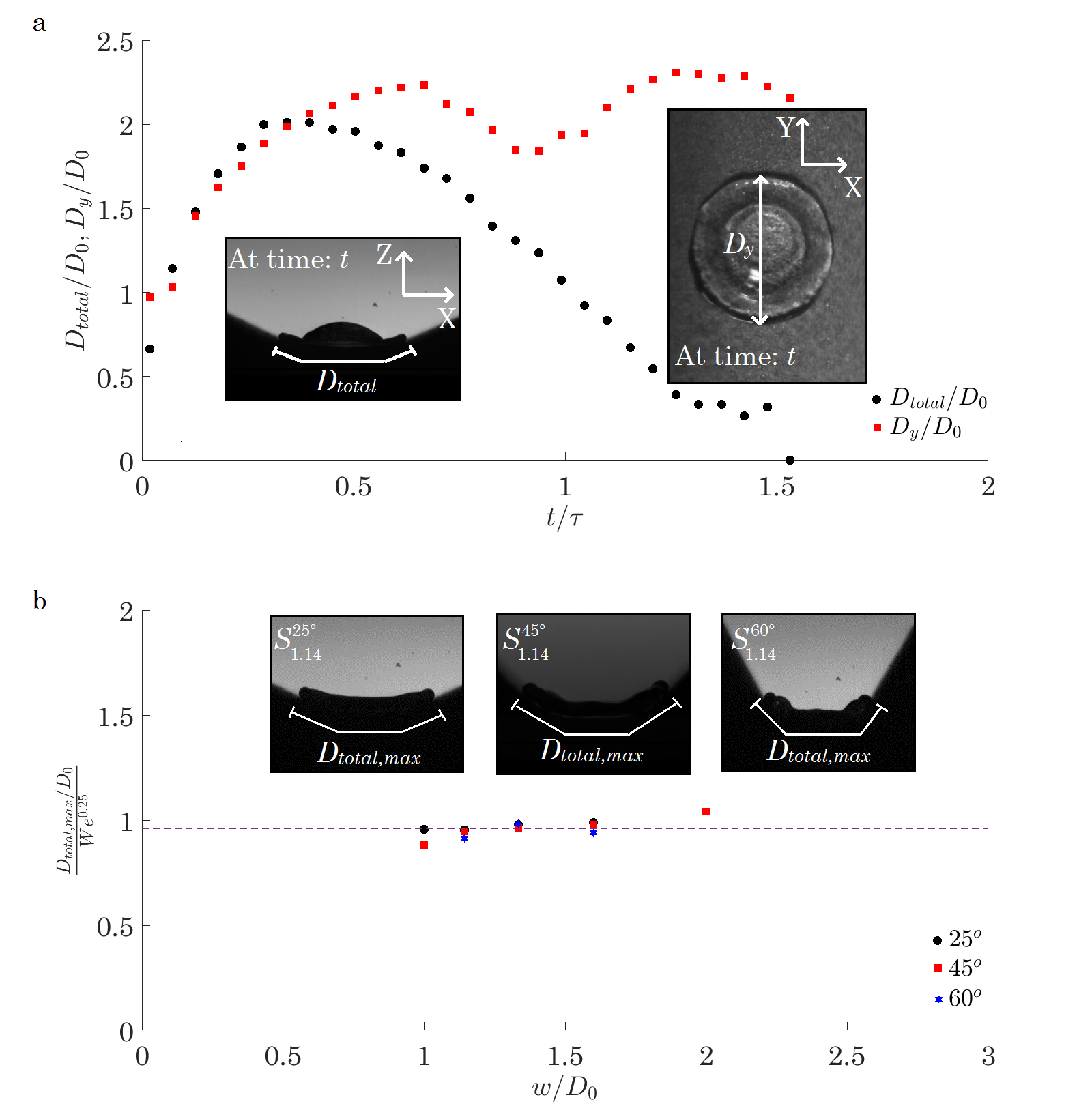}}
  \caption{\textbf{(a)} Plot showing the spread length $D_{total}$ and $D_y$ with time $t$ for $S_{1.14}^{45\degree}$ configuration. The lengths are non-dimensionalised with $D_0$ and the time with $\tau$. The images in the inset show $D_{total}$ and $D_y$ at any time instant $t$. \textbf{(b)} Plot showing $\frac{D_{total, max}/D_0}{We^{0.25}}$ with $w/D_0$ for $w \geqslant D_0$. The value of $D_{total, max}/D_0$ compares well with the scaling of \cite{CLANET2004} and is constant for all the configurations in this regime. Images in the inset show the $D_{total,max}$ for $S_{1.14}$ configurations.}
\label{new_spread}
\end{figure}

To investigate the spreading dynamics of the impact for $w \geqslant D_0$ regime, we consider the spread in two directions namely X and Y (see Figure \ref{new_morphology} and insets of \ref{new_spread}a). The spread length along X is the projection of the actual spread on the surface as a portion of drop spreads on the inclined face. Hence, we consider the total spread length ($D_{total}$) parallel to the surface at any time $t$ (see inset of Figure \ref{new_spread}a). The spread along the valley at the same time instant $t$ is considered to be $D_y$. 

For brevity, we consider the configuration $S_{1.14}^{45\degree}$ only to show the time evolution of $D_{total}$ and $D_y$ (see Figure \ref{new_spread}a). The most important feature of the plot is that $D_{total}$ {\raise.17ex\hbox{$\scriptstyle\mathtt{\sim}$}} $D_y$ up to the time to reach maximum $D_{total}$. In fact, $D_{total, max}/D_0$ {\raise.17ex\hbox{$\scriptstyle\mathtt{\sim}$}} $We^{0.25}$, compares well with the scaling of \cite{CLANET2004} for all the $S_{w \geqslant D_0}^{\theta}$ configurations as indicated in Figure \ref{new_spread}b. This signifies that the spreading is isotropic even in the presence of the inclined face in the X direction. Figure \ref{new_spread}a also shows that the receding in the X-direction is rapid compared to the Y-direction. Furthermore, the drop lifts-off even before the recoiling in the Y-direction is complete. The anisotropy is initiated in the receding phase. The liquid starts to form globules near the base of the inclined face (see Figure \ref{new_morphology}b). This local accumulation of mass produces lobes as explained earlier. Once the recoiling of the largest lobe is complete, the drop detaches from the surface with a reduced contact time.

\section{Conclusion}\label{CON}
With a systematic experimental procedure, we have shown that the tangential component of momentum and non-circular footprint of an impacting drop onto SHS is not a requirement to produce anisotropic flows. Even a lamella deflection can trigger the flow asymmetry and eventual reduction in contact time. The spread in the inertial regime is isotropic, and the asymmetric flow is originated during the recoiling. The variation of contact time with the width of the flat region is continuous and rapid in this regime. This contrasts the discrete behaviour of contact time with the impact velocity on macro-ridges. 

\section{Supplementary material}
The videos pertaining to a selected impact configurations ($S_{w/D_0}^{\theta}$) are provided.
\begin{itemize}
\item \textbf{Movie 1}: Drop impact movie with configuration $S_{0.17}^{45\degree}$ and $We = 19.8$.  
\item \textbf{Movie 2}: Drop impact movie with configuration $S_{0.5}^{45\degree}$ and $We = 21.4$.
\item \textbf{Movie 3}: Drop impact movie with configuration $S_{1}^{45\degree}$ and $We = 22.2$.
\item \textbf{Movie 4}: Drop impact movie with configuration $S_{1.14}^{45\degree}$ and $We = 20.5$.
\item \textbf{Movie 5}: Drop impact movie with configuration $S_{1.6}^{45\degree}$ and $We = 22.3$.
\item \textbf{Movie 6}: Drop impact movie with configuration $S_{2.7}^{45\degree}$ and $We = 20.7$.
\item \textbf{Movie 7}: Drop impact movie with configuration $S_{1.14}^{25\degree}$ and $We = 21.6$.
\item \textbf{Movie 8}: Drop impact movie with configuration $S_{1.6}^{25\degree}$ and $We = 22.8$.
\item \textbf{Movie 9}: Drop impact movie with configuration $S_{1.14}^{60\degree}$ and $We = 22.7$.
\item \textbf{Movie 10}: Drop impact movie with configuration $S_{1.6}^{60\degree}$ and $We = 21.4$.
\end{itemize}
\bibliographystyle{jfm}
\bibliography{nc}

\begin{thebibliography}{23}
\expandafter\ifx\csname natexlab\endcsname\relax\def\natexlab#1{#1}\fi
\def\au#1{#1} \def\ed#1{#1} \def\yr#1{#1}\def\at#1{#1}\def\jt#1{\textit{#1}}
  \def\bt#1{#1}\def\bvol#1{\textbf{#1}} \def\vol#1{#1} \def\pg#1{#1}
  \def\publ#1{#1}\def\arxiv#1{#1}\def\org#1{#1}\def\st#1{\textit{#1}}

\bibitem[Barthlott \& Neinhuis(1997)]{Barthlott1997}
{\sc \au{Barthlott, W.} \& \au{Neinhuis, C.}} \yr{1997}  \at{{Purity of the
  sacred lotus, or escape from contamination in biological surfaces}}.
  \jt{Planta}  \bvol{202}~(1),  \pg{1--8},  \arxiv{arXiv: arXiv:1011.1669v3}.

\bibitem[Bird {\em et~al.\/}(2013)Bird, Dhiman, Kwon \& Varanasi]{Bird2013}
{\sc \au{Bird, J~C}, \au{Dhiman, R}, \au{Kwon, H~M} \& \au{Varanasi, K~K}}
  \yr{2013}  \at{{Reducing the contact time of a bouncing drop}}.  \jt{Nature}
  \bvol{503}~(7476),  \pg{385--388}.

\bibitem[Cao {\em et~al.\/}(2009)Cao, Jones, Sikka, Wu \& Gao]{Cao2009}
{\sc \au{Cao, Liangliang}, \au{Jones, Andrew~K.}, \au{Sikka, Vinod~K.}, \au{Wu,
  Jianzhong} \& \au{Gao, Di}} \yr{2009}  \at{{Anti-Icing superhydrophobic
  coatings}}.  \jt{Langmuir}  \bvol{25}~(21),  \pg{12444--12448}.

\bibitem[Cassie \& Baxter(1944)]{Cassie1944}
{\sc \au{Cassie, A. B.~D.} \& \au{Baxter, S.}} \yr{1944}  \at{Wettability of
  porous surfaces}.  \jt{Trans. Faraday Soc.}  \bvol{40},  \pg{546--551}.

\bibitem[Chandra \& Avedisian(1991)]{Chandra13}
{\sc \au{Chandra, S.} \& \au{Avedisian, C.~T.}} \yr{1991}  \at{On the collision
  of a droplet with a solid surface}.  \jt{Proceedings of the Royal Society of
  London A: Mathematical, Physical and Engineering Sciences}
  \bvol{432}~(1884),  \pg{13--41}.

\bibitem[Clanet {\em et~al.\/}(2004)Clanet, B{\'{e}}guin, Richard \&
  Qu{\'{e}}r{\'{e}}]{CLANET2004}
{\sc \au{Clanet, Christophe}, \au{B{\'{e}}guin, C{\'{e}}dric}, \au{Richard,
  Denis} \& \au{Qu{\'{e}}r{\'{e}}, David}} \yr{2004}  \at{{Maximal deformation
  of an impacting drop}}.  \jt{Journal of Fluid Mechanics}  \bvol{517},
  \pg{199--208}.

\bibitem[Gauthier {\em et~al.\/}(2015)Gauthier, Symon, Clanet \&
  Qu{\'{e}}r{\'{e}}]{Gauthier2015}
{\sc \au{Gauthier, Ana{\"{i}}s}, \au{Symon, Sean}, \au{Clanet, Christophe} \&
  \au{Qu{\'{e}}r{\'{e}}, David}} \yr{2015}  \at{{Water impacting on
  superhydrophobic macrotextures}}.  \jt{Nature Communications}  \bvol{6},
  \pg{8001}.

\bibitem[Jung {\em et~al.\/}(2011)Jung, Dorrestijn, Raps, Das, Megaridis \&
  Poulikakos]{Jung2011}
{\sc \au{Jung, Stefan}, \au{Dorrestijn, Marko}, \au{Raps, Dominik}, \au{Das,
  Arindam}, \au{Megaridis, Constantine~M.} \& \au{Poulikakos, Dimos}} \yr{2011}
   \at{{Are superhydrophobic surfaces best for icephobicity?}}  \jt{Langmuir}
  \bvol{27}~(6),  \pg{3059--3066}.

\bibitem[Jung {\em et~al.\/}(2012)Jung, Tiwari, Doan \& Poulikakos]{Jung2012}
{\sc \au{Jung, Stefan}, \au{Tiwari, Manish~K}, \au{Doan, N~Vuong} \&
  \au{Poulikakos, Dimos}} \yr{2012}  \at{{Mechanism of supercooled droplet
  freezing on surfaces.}}  \jt{Nature communications}  \bvol{3},  \pg{615}.

\bibitem[Kulinich \& Farzaneh(2009)]{Kulinich2009}
{\sc \au{Kulinich, S.~A.} \& \au{Farzaneh, M.}} \yr{2009}  \at{{How wetting
  hysteresis influences ice adhesion strength on superhydrophobic surfaces}}.
  \jt{Langmuir}  \bvol{25}~(16),  \pg{8854--8856}.

\bibitem[Lafuma \& Qu{\'{e}}r{\'{e}}(2003)]{Lafuma2003}
{\sc \au{Lafuma, Aur{\'{e}}lie} \& \au{Qu{\'{e}}r{\'{e}}, David}} \yr{2003}
  \at{{Superhydrophobic states.}}  \jt{Nature materials}  \bvol{2}~(7),
  \pg{457--60}.

\bibitem[Liu {\em et~al.\/}(2015)Liu, Andrew, Li, Yeomans \& Wang]{Liu2015a}
{\sc \au{Liu, Yahua}, \au{Andrew, Matthew}, \au{Li, Jing}, \au{Yeomans,
  Julia~M} \& \au{Wang, Zuankai}} \yr{2015}  \at{{Symmetry-breaking in drop
  bouncing on curved surfaces}}.  \jt{Nature Communications}  \bvol{6},
  \pg{1--8},  \arxiv{arXiv: 1511.00064}.

\bibitem[Liu {\em et~al.\/}(2014)Liu, Moevius, Xu, Qian, Yeomans \&
  Wang]{Liu2014}
{\sc \au{Liu, Yahua}, \au{Moevius, Lisa}, \au{Xu, Xinpeng}, \au{Qian,
  Tiezheng}, \au{Yeomans, Julia~M} \& \au{Wang, Zuankai}} \yr{2014}
  \at{{Pancake bouncing on superhydrophobic surfaces}}.  \jt{Nature Physics}
  \bvol{10}~(June),  \pg{515--519},  \arxiv{arXiv: 1406.3203}.

\bibitem[Marmur(2004)]{Marmur2004}
{\sc \au{Marmur, Abraham}} \yr{2004}  \at{{The Lotus Effect :
  Superhydrophobicity and Metastability}}.  \jt{Langmuir} ~(20),
  \pg{3517--3519}.

\bibitem[Meuler {\em et~al.\/}(2010)Meuler, Smith, Varanasi, Mabry, McKinley \&
  Cohen]{Meuler2010}
{\sc \au{Meuler, Adam~J.}, \au{Smith, J.~David}, \au{Varanasi, Kripa~K.},
  \au{Mabry, Joseph~M.}, \au{McKinley, Gareth~H.} \& \au{Cohen, Robert~E.}}
  \yr{2010}  \at{{Relationships between water wettability and ice adhesion}}.
  \jt{ACS Applied Materials and Interfaces}  \bvol{2}~(11),  \pg{3100--3110}.

\bibitem[Onda {\em et~al.\/}(1996)Onda, Shibuichi, Satoh \&
  Tsujii]{Shibuichi1996}
{\sc \au{Onda, T.}, \au{Shibuichi, S.}, \au{Satoh, N.} \& \au{Tsujii, K.}}
  \yr{1996}  \at{Super-water-repellent fractal surfaces}.  \jt{Langmuir}
  \bvol{12}~(9),  \pg{2125--2127},  \arxiv{arXiv:
  http://dx.doi.org/10.1021/la950418o}.

\bibitem[Quere \& Reyssat(2008)]{Quere2008}
{\sc \au{Quere, D.} \& \au{Reyssat, M.}} \yr{2008}  \at{{Non-adhesive lotus and
  other hydrophobic materials}}.  \jt{Philosophical Transactions of the Royal
  Society A: Mathematical, Physical and Engineering Sciences}
  \bvol{366}~(1870),  \pg{1539--1556}.

\bibitem[Regulagadda {\em et~al.\/}(2017)Regulagadda, Bakshi \&
  Das]{kartik2017}
{\sc \au{Regulagadda, Kartik}, \au{Bakshi, Shamit} \& \au{Das, Sarit~Kumar}}
  \yr{2017}  \at{Morphology of drop impact on a superhydrophobic surface with
  macro-structures}.  \jt{Physics of Fluids}  \bvol{29}~(8),  \pg{082104},
  \arxiv{arXiv: http://dx.doi.org/10.1063/1.4997266}.

\bibitem[Richard {\em et~al.\/}(2002)Richard, Clanet \&
  Qu{\'{e}}r{\'{e}}]{Richard2002}
{\sc \au{Richard, Denis}, \au{Clanet, Christophe} \& \au{Qu{\'{e}}r{\'{e}},
  David}} \yr{2002}  \at{{Contact time of a bouncing drop.}}  \jt{Nature}
  \bvol{417}~(6891),  \pg{811}.

\bibitem[Tran {\em et~al.\/}(2012)Tran, Staat, Prosperetti, Sun \&
  Lohse]{Tran2012}
{\sc \au{Tran, Tuan}, \au{Staat, Hendrik J.~J.}, \au{Prosperetti, Andrea},
  \au{Sun, Chao} \& \au{Lohse, Detlef}} \yr{2012}  \at{{Drop Impact on
  Superheated Surfaces}}.  \jt{Physical Review Letters}  \bvol{108}~(3),
  \pg{036101}.

\bibitem[Weisensee {\em et~al.\/}(2016)Weisensee, Tian, Miljkovic \&
  King]{Weisensee2016}
{\sc \au{Weisensee, Patricia~B.}, \au{Tian, Junjiao}, \au{Miljkovic, Nenad} \&
  \au{King, William~P.}} \yr{2016}  \at{{Water droplet impact on elastic
  superhydrophobic surfaces}}.  \jt{Scientific Reports}  \bvol{6}~(July),
  \pg{30328}.

\bibitem[Wenzel(1936)]{Wenzel1936}
{\sc \au{Wenzel, Robert~N}} \yr{1936}  \at{{Resistance of solid surfaces to
  wetting by water.}}  \jt{Journal of Industrial and Engineering Chemistry
  (Washington, D. C.)}  \bvol{28},  \pg{988--994}.

\bibitem[Worthington(1876)]{Worthington01011876}
{\sc \au{Worthington, A.~M.}} \yr{1876}  \at{On the forms assumed by drops of
  liquids falling vertically on a horizontal plate}.  \jt{Proceedings of the
  Royal Society of London}  \bvol{25}~(171-178),  \pg{261--272}.

\end{thebibliography}

\end{document}